%% file: sigir2024_dense_query_tar.tex
\documentclass[sigconf]{acmart}
\usepackage{color} 
\usepackage{xspace} 
\usepackage{balance}
\usepackage{graphicx}

\usepackage{courier}
\usepackage{amsmath}
\usepackage{soul}
\usepackage{color}

\usepackage{enumerate}
\usepackage[inline]{enumitem}

\usepackage{booktabs}
\usepackage{multirow}
\usepackage{adjustbox}
\usepackage{dcolumn}

\usepackage{caption}
\usepackage{subcaption}
\usepackage{multirow}
\usepackage{wrapfig}
\usepackage{pifont}
\usepackage{fontawesome5}

\usepackage{tikz}
\usetikzlibrary{positioning,fit,backgrounds}
\AtBeginDocument{%
	\providecommand\BibTeX{{%
			\normalfont B\kern-0.5em{\scshape i\kern-0.25em b}\kern-0.8em\TeX}}}

\copyrightyear{2024}
\acmYear{2024}
\setcopyright{licensedothergov}\acmConference[SIGIR '24]{Proceedings of the 47th International ACM SIGIR Conference on Research and Development in Information Retrieval}{July 14--18, 2024}{Washington, DC, USA}
\acmBooktitle{Proceedings of the 47th International ACM SIGIR Conference on Research and Development in Information Retrieval (SIGIR '24), July 14--18, 2024, Washington, DC, USA}
\acmDOI{10.1145/3626772.3657921}
\acmISBN{979-8-4007-0431-4/24/07}

\begin{document}
	
	\title{Dense Retrieval with Continuous Explicit Feedback for Systematic Review Screening Prioritisation}
		
	\author{Xinyu Mao}
	\affiliation{%
		\institution{\parbox{5cm}{The University of Queensland}}
		\city{Brisbane}
		\country{Australia}
	}
	\email{xinyu.mao@uq.edu.au}
	
	\author{Shengyao Zhuang}
	\affiliation{%
		\institution{CSIRO}
		\city{Brisbane}
		\country{Australia}
	}
	\email{shengyao.zhuang@csiro.au}
	
	\author{Bevan Koopman}
	\affiliation{%
		\institution{CSIRO}
		\city{Brisbane}
		\country{Australia}
	}
	\email{bevan.koopman@csiro.au}
	
	\author{Guido Zuccon}
	\affiliation{%
		\institution{\parbox{5cm}{The University of Queensland}}
		\city{Brisbane}
		\country{Australia}
	}
	\email{g.zuccon@uq.edu.au}

\renewcommand{\shortauthors}{Xinyu Mao, Shengyao Zhuang, Bevan Koopman, \& Guido Zuccon}

\begin{abstract}

The goal of screening prioritisation in systematic reviews is to identify relevant documents with high recall and rank them in early positions for review. 
This saves reviewing effort if paired with a stopping criterion, 
and speeds up review completion if performed alongside downstream tasks. 
Recent studies have shown that neural models have good potential on this task, but their time-consuming fine-tuning and inference discourage their widespread use for screening prioritisation. 
In this paper, we propose an alternative approach that still relies on neural models, but leverages dense representations and relevance feedback to enhance screening prioritisation, without the need for costly model fine-tuning and inference. 
This method exploits continuous relevance feedback from reviewers during document screening to efficiently update the dense query representation, which is then applied to rank the remaining documents to be screened. 
We evaluate this approach across the CLEF TAR datasets for this task. Results suggest that the investigated dense query-driven approach is more efficient than directly using neural models and shows promising effectiveness compared to previous methods developed on the considered datasets. Our code is available at \url{https://github.com/ielab/dense-screening-feedback}.

\end{abstract}



\maketitle
	
\raggedbottom
\input{introduction}
\input{methods}
\input{experiment_settings}

\input{results}

\input{conclusions}

\section*{Acknowledgment}
Xinyu Mao is supported by the Australian Research Council Discovery Project DP210104043. 

%
%
%

\graphicspath{ {./figures/} }

\bibliographystyle{ACM-Reference-Format}
\balance
\bibliography{bibliography}


\end{document}

%% file: introduction.tex
\section{Introduction}

A medical systematic review aims to answer a specific medical question by collecting and appraising relevant studies as evidence. These reviews often require screening. Screening involves assessing documents for inclusion or exclusion in the review. A large number of retrieved documents, a task that is labour-intensive and time-consuming~\cite{mcgowan2005systematic,michelson2019significant,borah2017analysis}. Screening is done in two phases: 1) document titles and abstracts are assessed first for relevance; and 2) full-text articles from phase 1 are assessed for relevance \cite{higgins2019cochrane}. These two phases are conceptually executed sequentially; i.e., after filtering out a large number of irrelevant documents based on title and abstract, full text screening is started on the remaining documents.

Screening prioritisation can reduce the time needed to complete the review and becomes a critical task for systematic review automation methods, such as technology-assisted review (TAR) \cite{lee2018seed}.  In screening prioritisation, documents are ranked, ideally in a way that places relevant documents above non-relevant ones, allowing them to be screened first.
 This enables downstream tasks such as full-text screening to begin as soon as a relevant document is found. If all relevant documents are encountered immediately, then downstream tasks might be completed concurrently as other researchers continue screening the remaining irrelevant documents,
  thus shortening the time to write the review \cite{omara-eves2015using}. Screening prioritisation can also save overall effort (and thus cost) when early stopping is applied to halt screening after top-k documents are reviewed, thereby avoiding the exhaustive screening of all documents \cite{wallace2010semi, cormack2016engineering, yang2021heuristic}.

There are two main approaches to screening prioritisation~\cite{wangs2022neural}: 
\begin{enumerate*}
	\item \textbf{Query-based:} The query associated to the systematic review is used to rank documents, often using sparse representations such as BM25 \cite{di2017interactive, alharbi2019ranking} and TF-IDF \cite{alharbi2017ranking, scells2017qut}. Queries are often formed from the working title of the review \cite{alharbi2017ranking, alharbi2018retrieving, alharbi2019ranking}, its research questions \cite{scells2017qut}, or the Boolean query used to retrieve the documents to screen \cite{alharbi2017ranking, alharbi2018retrieving, alharbi2019ranking,  chen2017ecnu}.

	\item \textbf{Model-based:} A classification model is trained to discriminate documents into inclusion or exclusion classes. Most work on this task has focused on the use of traditional machine learning models such as SVM \cite{yu2017data, anagnostou2017hybridranksvm} and logistic regression \cite{wu2018ecnu, norman2017limsi, li2019automatic}. Recent studies have shown the potential of BERT-based models for this task \cite{wangs2022neural, yang2022goldilocks, molinari2022transferring}. 
	Many techniques are also leveraged together with these models, such as active learning \cite{wallace2010semi, yu2017data, li2019automatic, grossman2017automatic, norman2018limsi, cormack2017technology, cormack2018technology}, and relevance feedback \cite{kalphov2017sis, yu2017data, anagnostou2017hybridranksvm, cormack2017technology, cormack2018technology}, which is also seen in combination with query-based methods \cite{alharbi2018retrieving, alharbi2019ranking,di2017interactive, di2018interactive, di2019distributed}.
\end{enumerate*}  

Model-based approaches in TAR often incorporate \textit{active learning}, such as CAL \cite{cormack2014evaluation}, where a SVM classifier is trained with continuous feedback from top-ranked documents. Specifically, linear classifiers require relevant seed documents to initiate training. However, in systematic reviews, seed studies are served for query formulation and are not always relevant to the topic \cite{wang2022little}, which may harm retrieval effectiveness when initiating the active learning workflow in a real systematic review setting. Additionally, the size of the training set also increases due to the active learning mechanism, though the classifiers can still benefit by combining knowledge from retrieval (such as search keywords) regardless of the varying size of the training set \cite{yang2019text}.
Recently, a BERT-based TAR workflow \cite{yang2022goldilocks} has shown a promising trend, delivering higher effectiveness when prioritising documents for screening in systematic reviews, especially if applied with a domain-specific backbone without any further pre-training~\cite{mao2024reproducibility}. A drawback is the method incurs higher costs in time and computation compared to linear models~\cite{mao2024reproducibility}. 
Query-based approaches have mainly been investigated in static, once-off rankings. 
These methods could be promising in a continuous feedback setting, especially if they can deliver similar effectiveness to model-based approaches but at a lower computational cost.

In this paper, we follow the query-based approach but adopt neural encoder models, such as BERT-based dense retrievers, to perform screening prioritisation that exploits the human reviewer's iterative feedback. While BERT-based rankers have been shown highly effective for screening prioritisation, no relevance feedback mechanism has been investigated~\cite{wangs2022neural}. 
Methods for using relevance feedback in combination with BERT-based rankers have been devised for ad-hoc retrieval~\cite{yu2021improving,wang2020pseudo,zheng2021contextualized}; however, their limitations include  (1) only considering pseudo-relevance feedback, (2) only implementing once-off setting (i.e., the feedback mechanism is used only once to produce a ranking), (3) most methods re-train the ranker after feedback (through additional fine-tuning) \cite{yu2021improving,baumgartner2022incorporating,li2022improving}. An exception is Li et al.~\cite{li2023pseudo}, which we adapt here to screening prioritisation.
The task of screening prioritisation differs in that the feedback is explicit, and is provided in a continuous manner as reviewers perform screening.
Typically, adapting current BERT-based rankers for screening prioritisation with relevance feedback is computationally expensive due to iterative re-fine-tuning. We therefore explore an unexamined alternative: using dense retrievers with an efficient strategy for leveraging feedback~\cite{li2023pseudo}. Our results show that this approach is not only efficient but also matches or exceeds the performance of specialised methods for screening prioritisation.

%% file: methods.tex
\section{Dense Retrieval with Continuous Explicit Feedback} 

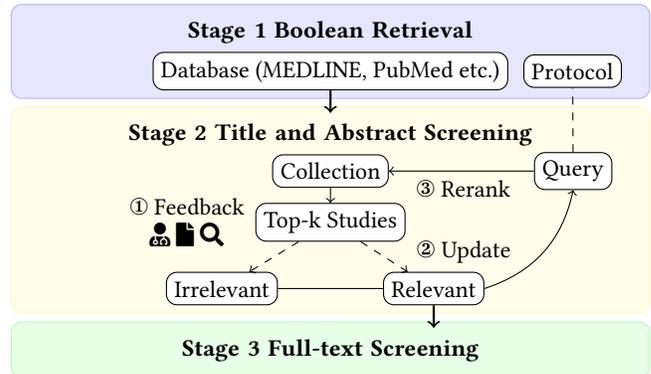
\begin{figure}[t]
	\centering
\pgfdeclarelayer{background}
\pgfdeclarelayer{foreground}
\pgfsetlayers{background,main,foreground}
\begin{tikzpicture}

\node[align=center,text width=.95\columnwidth] (stage1-title) {\bf Stage 1 Boolean Retrieval};
\node[draw,rounded corners, below=0mm of stage1-title,fill=white] (stage1-database) {Database (MEDLINE, PubMed etc.)};
\node[draw,rounded corners,right=2mm of stage1-database,fill=white] (stage1-protocol) {Protocol};

\begin{pgfonlayer}{background}
\node[draw,rounded corners,text width=\columnwidth,align=center,color=blue!30,fill=blue!10,fit=(stage1-title)(stage1-database)] (stage1) {};
\end{pgfonlayer}

\node[align=center,below=11mm of stage1.south |- stage1-title,text width=.95\columnwidth] (stage2-title) {\bf Stage 2 Title and Abstract Screening};
\node[draw,rounded corners, below=0mm of stage2-title,fill=white] (stage2-collection) {Collection};
\node[draw,rounded corners, right=19.3mm of stage2-collection,fill=white,label={[anchor=north east]180,xshift=-2mm:\ding{174}~Rerank}] (stage2-query) {Query};
\node[draw,rounded corners, below=2mm of stage2-collection,fill=white,label={[align=center]180:\parbox{16mm}{\centering\ding{172}~Feedback\\\faUserMd~\faFile~\faSearch}}] (stage2-topk) {Top-k Studies};	
\node[draw,rounded corners, below left=4mm and -3mm of stage2-topk,fill=white] (stage2-irrelevant) {Irrelevant};
\node[draw,rounded corners, below right=4mm and -3mm of stage2-topk,fill=white,label={90,xshift=4mm:\ding{173}~Update}] (stage2-relevant) {Relevant};

\begin{pgfonlayer}{background}
\node[draw,rounded corners,text width=\columnwidth,align=center,color=yellow!30,fill=yellow!10,fit=(stage2-title)(stage2-collection)(stage2-topk)(stage2-irrelevant)(stage2-relevant)] (stage2) {};
\end{pgfonlayer}

\node[below=26mm of stage2.south |- stage2-title,text width=.95\columnwidth,align=center] (stage3) {\bf Stage 3 Full-text Screening};
\begin{pgfonlayer}{background}
	\node[draw,rounded corners,text width=\columnwidth,align=center,color=green!30,fill=green!10,fit=(stage3)] {};
\end{pgfonlayer}

\draw[->,thick] (stage1-database.south) to +(0,-.3);
\draw[-, dashed] (stage2-query) to (stage1-protocol);
\draw[->] (stage2-query) to (stage2-collection);
\draw[->] (stage2-collection) to (stage2-topk);
\draw[->,dashed] (stage2-topk) to (stage2-irrelevant);
\draw[->,dashed] (stage2-topk) to (stage2-relevant);
\draw[-] (stage2-irrelevant) to (stage2-relevant);
\draw[->,bend right] (stage2-relevant.east) to (stage2-query.south);
\draw[->,thick] (stage2-relevant.south) to +(0,-.3);
\end{tikzpicture}
	\caption{Screening prioritisation with feedback.}
	\label{fig:1}
\end{figure}

Figure~\ref{fig:1} is an overview of our proposed dense retrieval framework for screening prioritisation with continuous feedback.
In Stage 1, users formulate Boolean queries to retrieve documents from databases such as PubMed. These documents form the pool for title and abstract screening in Stage 2. A protocol, defining key questions of the review and the inclusion/exclusion criteria, is also available at Stage 1.
We then propose to utilise topic-related information from the protocol as the query for dense retrieval against the pool, and then select top-k documents for user examination. Users assess these documents as relevant or non-relevant using titles and abstracts. These binary judgments then serve as explicit relevance feedback, updating the query with Rocchio's algorithm on the dense representations~\cite{li2023pseudo}:
$
\overrightarrow{q}_{\text{update}}  =  \alpha* \overrightarrow{q} + \beta * \text{avg}(\overrightarrow{d}^{+}_{1}, ... ,\overrightarrow{d}^{+}_{m}) - \gamma * \text{avg}(\overrightarrow{d}^{-}_{1}, ... ,\overrightarrow{d}^{-}_{n})
$
where $\overrightarrow{d}^{+}$ and $\overrightarrow{d}^{-}$ are relevant and non-relevant dense representations (as judged by reviewers) and $\overrightarrow{q}$ is the dense query. $\alpha$, $\beta$, and $\gamma$ are the Rocchio weights of the previous query, relevant documents, and non-relevant documents' dense representation, respectively. The refined dense query representation is used to obtain a new ranking over the remaining documents to be screened. This process is repeated until all documents are reviewed or a stopping point is reached.
Documents identified as relevant at each iteration can be directly passed to the downstream tasks (e.g., full-text screening), accelerating the systematic review process.
The effectiveness and efficiency of Rocchio's algorithm on dense representation of queries and documents have been studied for once-off, static creation of a ranking based on pseudo-relevance feedback and for ad-hoc retrieval; however, its use in continuous feedback and for systematic review screening has not been explored.

%% file: experiment_settings.tex
\section{Experiment Setup}

\begin{table*}[!htb]
    \begin{minipage}{.69\linewidth}
      \caption{Initial ranking refers to the results obtained before screening prioritisation with continuous feedback. ($\alpha$, $\beta$, $\gamma$) presents different feedback settings. $\dagger$ indicates statistical significant difference between dense retrievers and BM25+RM3 for initial rankings, while * between no feedback and continuous feedback. Statistical significance has been computed using paired t-test with Bonferroni correction, $p<0.05$.}
	\label{tab1:clef-all}
      \centering
        \resizebox{1\textwidth}{!}{%
		\begin{tabular}{cc|cccccccccc}
			\toprule
			\multirow{2}[4]{*}{\textbf{Collections}} & \multicolumn{1}{c}{\multirow{2}[4]{*}{\textbf{Methods}}} & \multicolumn{2}{c}{\textbf{initial ranking}} & \multicolumn{2}{c}{\textbf{(1,1,1)}} & \multicolumn{2}{c}{\textbf{ (1,0.5,0.5)}} & \multicolumn{2}{c}{\textbf{(1,0.8,0.2)}} & \multicolumn{2}{c}{\textbf{(1,1,0)}} \\\cmidrule{3-12}          & \multicolumn{1}{c}{} & \textbf{AP} & \textbf{Last Rel} & \textbf{AP} & \textbf{Last Rel} & \textbf{AP} & \textbf{Last Rel} & \textbf{AP} & \textbf{Last Rel} & \textbf{AP} & \textbf{Last Rel} \\    \midrule    \multirow{6}[8]{*}{CLEF 17} &\textbf{BM25+RM3}& .1439 & 2228  & -     & -     & -     & -     & -     & -     & -     & - \\\cmidrule{2-12}          & \textbf{coCondenser}& .2064 & \multicolumn{1}{c|}{2477} & .2335 & \multicolumn{1}{c|}{*3770} & *.2357 & \multicolumn{1}{c|}{*3734} & *.2404 & \multicolumn{1}{c|}{2110} & \textbf{*.2370} & \textbf{1979} \\\cmidrule{2-12}          & \textbf{BioLinkBERT} & .2402 & \multicolumn{1}{c|}{\textbf{1917}} & \textbf{*.2738} & \multicolumn{1}{c|}{*3620} & *.2707 & \multicolumn{1}{c|}{*3538} & .2543 & \multicolumn{1}{c|}{2087} & .2431 & 2314 \\          & \textbf{PubMedBERT} & .1597 & \multicolumn{1}{c|}{\textbf{2669}} & \textbf{*.1914} & \multicolumn{1}{c|}{3560} & *.1830 & \multicolumn{1}{c|}{3598} &*.1227 & \multicolumn{1}{c|}{2850} & *.1144 & *3423 \\          & \textbf{BioBERT} & *.1083 & \multicolumn{1}{c|}{$\dagger$3278} & .1161 & \multicolumn{1}{c|}{*3767} & \textbf{.1175} & \multicolumn{1}{c|}{*3777} & .1095 & \multicolumn{1}{c|}{3315} & .1076 & \textbf{3113} \\\cmidrule{2-12}          & \multicolumn{11}{c}{\textbf{auth.simple.run1 \cite{anagnostou2017hybridranksvm} }  AP: \textbf{.2970}  Last Rel: \textbf{2143}  }
			\\    \midrule    \multirow{6}[8]{*}{CLEF 18} & \textbf{BM25+RM3} & .1958 & 3276  & -     & -     & -     & -     & -     & -     & -     & - \\\cmidrule{2-12}          & \textbf{coCondenser}& $\dagger$.2798 & \multicolumn{1}{c|}{5291} & *.3420  & \multicolumn{1}{c|}{*4476} & \textbf{*.3472} & \multicolumn{1}{c|}{*4315} & *.3520 & \multicolumn{1}{c|}{\textbf{2590}} & *.3456 & 2609 \\\cmidrule{2-12}          & \textbf{BioLinkBERT} & $\dagger$.3601 & \multicolumn{1}{c|}{5047} & \textbf{*.4274} & \multicolumn{1}{c|}{*4027} & *.4213 & \multicolumn{1}{c|}{*4015} & .3816 & \multicolumn{1}{c|}{\textbf{2639}} & .3656 & 2954 \\          & \textbf{PubMedBERT} & $\dagger$.3284 & \multicolumn{1}{c|}{5673} & \textbf{*.3916} & \multicolumn{1}{c|}{3764} & *.3818 & \multicolumn{1}{c|}{3773} & *.2047 & \multicolumn{1}{c|}{\textbf{3304}} & *.1826 & *4236 \\          & \textbf{BioBERT} & .1748 & \multicolumn{1}{c|}{$\dagger$6335} & \textbf{*.2331} & \multicolumn{1}{c|}{4536} & *.2287 & \multicolumn{1}{c|}{4552} & .1783 & \multicolumn{1}{c|}{3839} & .1692 & \textbf{3701} \\\cmidrule{2-12}          & \multicolumn{11}{c}{\textbf{cnrs\_comb \cite{norman2018limsi}}  AP: \textbf{.3470} Last Rel: \textbf{2406}}
			\\    \midrule    \multirow{6}[8]{*}{CLEF 19 dta} & \textbf{BM25+RM3} & .1677 & 2633  & -     & -     & -     & -     & -     & -  & -     & - \\\cmidrule{2-12}          & \textbf{coCondenser}& .1904 & \multicolumn{1}{c|}{\textbf{1232}} & .2086 & \multicolumn{1}{c|}{2932} & .2158 & \multicolumn{1}{c|}{2893} & .2337 & \multicolumn{1}{c|}{2178} & \textbf{.2400} & 1288 \\\cmidrule{2-12}          & \textbf{BioLinkBERT} & .2239 & \multicolumn{1}{c|}{1255} & .2567 & \multicolumn{1}{c|}{2822} & .2529 & \multicolumn{1}{c|}{2759} & \textbf{.2568} & \multicolumn{1}{c|}{\textbf{876}} & .2534 & 1143 \\          & \textbf{PubMedBERT} & .2288 & \multicolumn{1}{c|}{\textbf{891}} & \textbf{.2536} & \multicolumn{1}{c|}{2670} & .2485 & \multicolumn{1}{c|}{2636} & .1914 & \multicolumn{1}{c|}{1256} & .1763 & 2000 \\          & \textbf{BioBERT} & .1723 & \multicolumn{1}{c|}{\textbf{1577}} & .1925 & \multicolumn{1}{c|}{2912} & .1927 & \multicolumn{1}{c|}{2857} & .1865 & \multicolumn{1}{c|}{1638} & \textbf{.1824} & 1722 \\   \midrule    \multirow{6}[8]{*}{CLEF 19 int.} & \textbf{BM25+RM3} & .1620  & 1178  & -     & -     & -     & -     & -     & -     & -     & - \\\cmidrule{2-12}          & \textbf{coCondenser}& $\dagger$.3459 & \multicolumn{1}{c|}{1172} & .3760  & \multicolumn{1}{c|}{1886} & .3725 & \multicolumn{1}{c|}{1872} & \textbf{*.3794} & \multicolumn{1}{c|}{\textbf{658}} & *.3747 & 830 \\\cmidrule{2-12}          & \textbf{BioLinkBERT} & $\dagger$.3876 & \multicolumn{1}{c|}{964} & \textbf{.4151} & \multicolumn{1}{c|}{1628} & .4087 & \multicolumn{1}{c|}{1734} & *.4088 & \multicolumn{1}{c|}{\textbf{707}} & *.4041 & 961 \\          & \textbf{PubMedBERT} & $\dagger$.2960  & \multicolumn{1}{c|}{\textbf{946}} & \textbf{*.3261} & \multicolumn{1}{c|}{1891} & *.3178 & \multicolumn{1}{c|}{1619} & *.2702 & \multicolumn{1}{c|}{1113} & *.2666 & 1416 \\          & \textbf{BioBERT} & .1251 & \multicolumn{1}{c|}{1713} & .1475 & \multicolumn{1}{c|}{1959} & \textbf{.1495} & \multicolumn{1}{c|}{1957} & {*.1367} & \multicolumn{1}{c|}{\textbf{1576}} & .1332 & 1579   \\
			\bottomrule
		\end{tabular}
	}
    \end{minipage}
    \hspace{.1cm}
    \begin{minipage}{.29\linewidth}
    \vspace{-18pt}
      \centering
        \caption{Comparison between our dense retrieval method and TAR active learning with relevance feedback for screening prioritisation. B stands for BioLinkBERT, L for logistic regression; title and pos for different seed settings; * for statistical significance computed as in Table 1.} 
	\label{tab:tar-compare}
        \resizebox{1\textwidth}{!}{%
		\begin{tabular}{cccc}
	\hline
	\multicolumn{4}{c}{20 iteration (cut-off @ 500)} \\ \hline
	\textbf{Collections} & \textbf{Runs} & \textbf{AP} & \textbf{Last Rel} \\ \hline
	\multirow{5}{*}{CLEF 17} & \multicolumn{1}{c|}{dense - B - best} & \textbf{.2660} & \textbf{241} \\ \cline{2-4} 
	& \multicolumn{1}{c|}{tar - B - title} & *.0674 & *358 \\
	& \multicolumn{1}{c|}{tar - B - pos} & \textbf{*.1431} & \textbf{*331} \\ \cline{2-4} 
	& \multicolumn{1}{c|}{tar - L - title} & .2240 & *322 \\
	& \multicolumn{1}{c|}{tar - L - pos} & \textbf{.2407} & \textbf{*318} \\ \hline
	\multirow{5}{*}{CLEF 18} & \multicolumn{1}{c|}{dense - B - best} & \textbf{.4071} & \textbf{269} \\ \cline{2-4} 
	& \multicolumn{1}{c|}{tar - B - title} & *.1292 & *378 \\
	& \multicolumn{1}{c|}{tar - B - pos} & \textbf{*.1971} & \textbf{*342} \\ \cline{2-4} 
	& \multicolumn{1}{c|}{tar - L - title} & *.2709 & *361 \\
	& \multicolumn{1}{c|}{tar - L - pos} & \textbf{.3005} & \textbf{334} \\ \hline
	\multirow{5}{*}{CLEF 19 dta} & \multicolumn{1}{c|}{dense - B - best} & \textbf{.2547} & \textbf{268} \\ \cline{2-4} 
	& \multicolumn{1}{c|}{tar - B - title} & *.1177 & \textbf{331} \\
	& \multicolumn{1}{c|}{tar - B - pos} & \textbf{.1681} & 356 \\ \cline{2-4} 
	& \multicolumn{1}{c|}{tar - L - title} & .1898 & 343 \\
	& \multicolumn{1}{c|}{tar - L - pos} &  \textbf{.2815} & \textbf{314} \\ \hline
	\multirow{5}{*}{CLEF 19 int.} & \multicolumn{1}{c|}{dense - B - best} & \textbf{.4100} & \textbf{175} \\ \cline{2-4} 
	& \multicolumn{1}{c|}{tar - B - title} & *.0937 & *281 \\
	& \multicolumn{1}{c|}{tar - B - pos} & \textbf{*.1410} & \textbf{*268} \\ \cline{2-4} 
	& \multicolumn{1}{c|}{tar - L - title} & *.2352 & *257 \\
	& \multicolumn{1}{c|}{tar - L - pos} & \textbf{*.2578} & *257 \\ \hline
\end{tabular}
}
    \end{minipage} 
\end{table*}

\textbf{Datasets.}
We rely on the CLEF-TAR 2017, 2018 and 2019 Subtask 2 datasets \cite{kanoulas2017clef,kanoulas2018clef,kanoulas2019clef} (abbreviated as CLEF 17-19). These datasets contain queries (topics), documents (title and abstract only) and relevance assessments associated with real systematic reviews.
We use the working title of the review as the query, which is available in the protocol file under each topic in the datasets.
Each dataset has one training set and one test set. We use the training set to sample positive and negative documents for training a dense retriever; while we use the test set for the iterative ranking and the evaluation.

\textbf{Dense Retriever Training and Retrieval.} For dense retrievers, we use domain-specific (BioBERT \cite{lee2020biobert}, PubMedBERT \cite{gu2021domain}, BiolinkBERT \cite{yasunaga2022linkbert}) and task-specific (coCondenser \cite{gao2021unsupervised})  as backbones. Models were trained using Tevatron \cite{gao2022tevatron}
 with 10 training passages  per topic (comprising 1 positive and 9 negatives) in a triplet loss $<\text{topic}, {d}^{+}, {d}^{-}>$ \cite{karpukhin2020dense}, on a single NVIDIA V100 with 32GB memory for 60 epochs. For dense retrieval with relevance feedback, we used Pyserini
\cite{lin2021pyserini}
  for retrieval  and FAISS~\cite{douze2024faiss}
   for encoding, and indexing the queries and corpus.
 Average runtime per collection was 2.6 minutes for training, 1.2 minutes for encoding/indexing, and 5 minutes per retrieval setting (model, Rocchio setting, feedback size), with time increasing for smaller feedback sizes.

\textbf{Relevance Feedback Settings.} Rocchio's algorithm has three parameters $\alpha$, $\beta$, $\gamma$, for which we have four settings: 
(1, 1, 1), (1, 0.8, 0.2), (1, 0.5, 0.5), (1, 1, 0). 
This allows us to explore the impact of including only positive feedback and varying degrees of negative feedback. 
Another parameter is the number of documents included in each feedback iteration, denoted as $k$, which is typically fixed throughout an experiment. We set $k=25$ in line with previous work in the systematic reviews field~\cite{singh2018improving}, but also explore the effects of varying $k$ among $\{5, 10, 15, 25, 50\}$ in one of our research questions.

\textbf{Baselines.} 
We first compare our method to BM25+RM3 (pseudo relevance feedback) for the effectiveness in initial rankings, where the corresponding dense retrievers do not use any feedback.
We then select the best run (AP) submitted to CLEF for the feedback settings, with feedback and without stopping strategies. Specifically: From CLEF 17, \textit{auth.simple.run1}~\cite{anagnostou2017hybridranksvm}, which was a hybrid classifier with LTR features iteratively trained with explicit feedback. 
From CLEF 18, \textit{cnrs\_comb}~\cite{norman2018limsi}, which was a neural network trained from a logistic regression and a CAL model, using task description as seed. 
CLEF 19 had no suitable runs to compare with.
We also compare against a recent BERT-based active learning workflow proposed for TAR~\cite{yang2022goldilocks}. We consider both neural and traditional linear classifiers for the TAR method. Specifically, we select BioLinkBERT~\cite{yasunaga2022linkbert} without further pre-training as suggested by~\cite{mao2024reproducibility} and logistic regression as in~\cite{yang2022goldilocks}. For a fair comparison, we apply the same recording approach to the TAR methods by concatenating the rank of feedback documents in each iteration as introduced below, which is different from the previous recording for TAR detailed in~\cite{mao2024reproducibility}. We test different seed document settings: the review title (title) as above and one relevant document (pos). We limit both the active learning and our feedback method to 20 iterations as running BERT-based active learning is computationally expensive, especially on large topics.
For this setting, we run experiments on an NVIDIA H100 (80GB).%

\textbf{Evaluation.}
We examine a continuous, iterative relevance feedback task where subsets of documents are progressively re-ranked. Specifically, if $n$ documents are ranked at iteration $i$, only $n-k$ are re-ranked at $i+1$, with $k$ being the feedback batch size. This results in $\lceil {n/k} \rceil$ rankings by the end. We unify these into a single ranking, maintaining the order from each iterative batch without revisions. This is distinct from some CLEF-TAR methods that may re-order already examined rank positions based on new relevance labels, potentially overestimating effectiveness.
Our task focuses on whether relevance feedback can benefit screening prioritisation. Therefore, we exhaust all candidate documents under each topic and record the reviewed documents in each iteration. We then measure on the concatenated list of reviewed documents (feedback) generated within the dense retrieval framework, instead of examining a new ranked list as with a reranker.
We use Average Precision (AP) and Last Relevant Found (Last Rel -- the position of the last retrieved relevant document) to measure the effectiveness of the ranking methods.
We also report the run time as the measure of efficiency.

%% file: results.tex
\section{Results}
In Table \ref{tab1:clef-all} we report the results obtained by BM25+RM3, the dense retrieval methods with/without feedback, and the best runs from CLEF when available. The results show that all dense retrievers except BioBERT outperform BM25+ RM3 in terms of AP when no feedback is considered, strengthening the use case of dense rankers in systematic review screening.
Interestingly, the task-specific dense retriever (coCondenser) sometimes outperforms domain-specific retrievers such as BioBERT, which use a biomedical BERT.
However, effectiveness differences are observed in Last Rel: methods that obtain high AP do not always obtain low Last Rel. For example, in CLEF 2018, despite significant AP gains by BioLinkBERT and PubMedBERT, these are not reflected in Last Rel where, compared to BM25+RM3, reviewers have to screen about 2000 more documents.

When examining results obtained with explicit relevance feedback, we observe that this always improves AP across all dense retrievers, given a suitable weight setting.
We then consider the impact of different weight combinations have on the dense relevance feedback; for this, we again analyse Table \ref{tab1:clef-all}. We identify that most of the top-performing results for the dense retrievers that use domain-specific backbones are obtained when the same weight is assigned to the query, relevant documents and non-relevant document representations (i.e. setting (1,1,1)). For coCondenser, instead, improvements are generally observed when non-relevant documents are given lower weight (i.e. (1,1,0) and (1,0.8,0.2)).

In Table~\ref{tab1:clef-all} we also contextualise the effectiveness of the examined relevance feedback technique with that of compatible runs submitted to the CLEF shared tasks. Direct comparison is difficult as we are not aware of the exact settings of feedback used by the CLEF runs (e.g., whether we use the same feedback size, or if examined documents are reordered once their relevance is observed). However, the results suggest that the feedback method studied here provides similar effectiveness, and warrants further comparative analysis with previous methods (for which code is often not released).
%
We then would like to see how the dense retrieval method, which does not involve iterative fine-tuning, compares to the popular TAR method, where a classifier is continuously trained. The TAR workflow consists of an active learning strategy and a classifier. Previous studies report higher effectiveness with a relevance feedback strategy~\cite{mao2024reproducibility}, that is, screening the top-k documents suggested by the model. We also follow this practice, as another common strategy, uncertainty sampling~\cite{salton1990improving} does not promote documents for higher relevance and is therefore not suitable for screening prioritisation. Another typical feature of the TAR approach is that it requires at least one relevant document as the seed to initialise. Typically, the seed documents are randomly sampled from the relevant document pool for experiments~\cite{yang2022goldilocks}. This practice may fit scenarios where certain related studies are identified prior to screening for a systematic review and can lead to higher effectiveness of the classifier. However, it is not fair to directly compare to the proposed dense retrieval method, where only topic-related information is used as the query. Additionally, to measure how these methods can prioritise relevant documents during the screening phase, we keep tracking those feedback sets and concatenate them into an overall ranking.

We report our results in Table \ref{tab:tar-compare}, where a cutoff is set at 20 feedback iterations (totalling 25 x 20 = 500 documents). For dense retrieval, we select BioLinkBERT as the backbone and report the best result from the Rocchio settings we use. For the TAR workflow, we examine BioLinkBERT and logistic regression as the classifiers. Generally, we find the linear model to be more effective compared to the BERT-based model for TAR, with a gap of at least 10\% in AP across all the CLEF collections, showing less difference in terms of Last Rel. When initiated with a relevant document, both methods show large improvements compared with using the title, which suggests TAR methods rely on a good seed to start. When turning to the dense retrieval method, however, it shows significantly higher effectiveness on both AP and Last Rel, except for in CLEF 19 dta, where TAR with logistic regression has 0.2815 in AP and 314 in Last Rel, whereas the dense retrieval method has 0.2547 and 268, respectively. This shows that the query-driven dense retrieval methods can effectively prioritise relevant documents with fewer iterations, with no prior reviewed relevant document involved.
In terms of efficiency, in the 20 iteration setting, the dense retrieval method takes $\approx$9 seconds per topic, which is competitive with that of TAR using logistic regression, which takes on average 5 seconds per topic. However, TAR with BioLinkBERT requires $\approx$10 minutes per topic (including fine-tuning and inference in each iteration).

 \begin{figure}[t]
	\centering
	\includegraphics[width=\columnwidth]{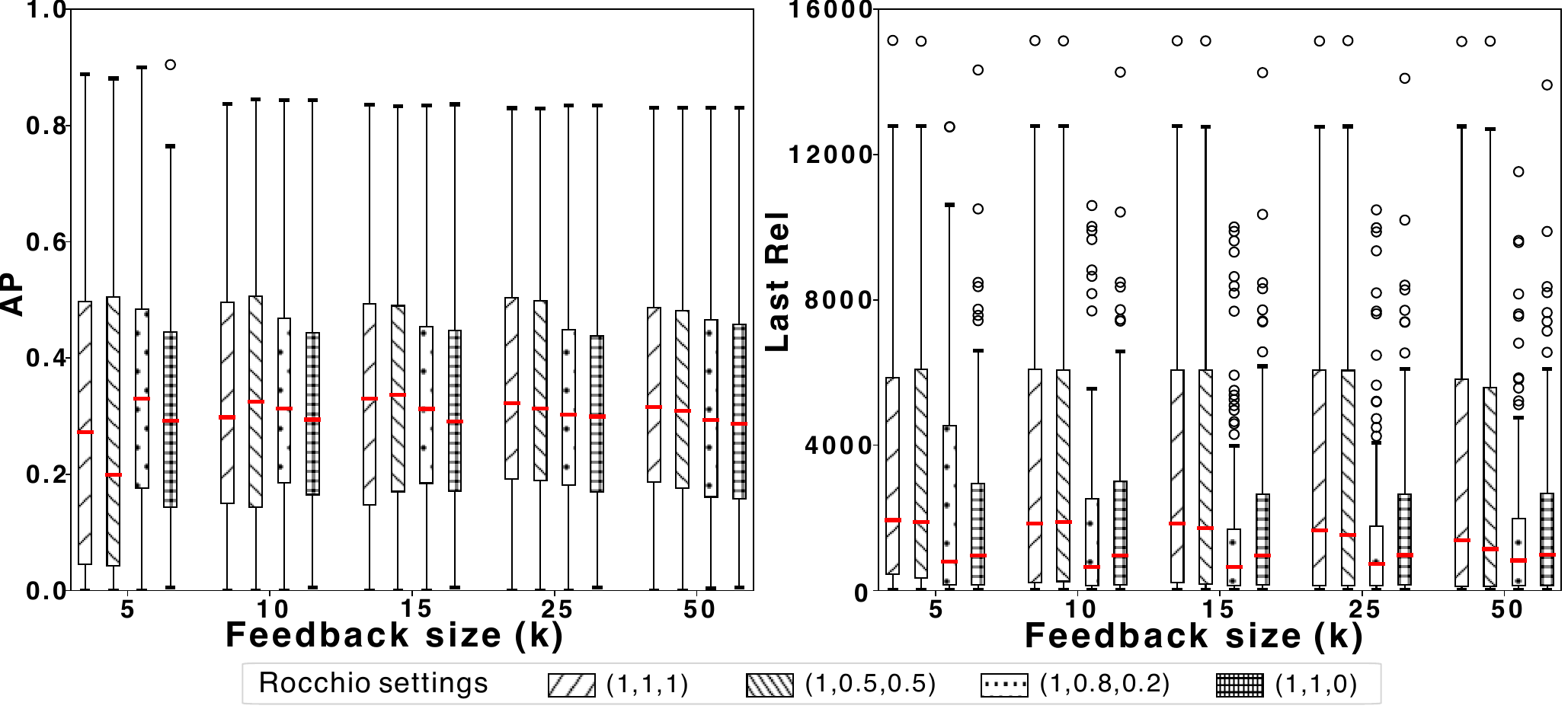}
	\caption{Distributions of Average Precision (left) and Last Relevant (right)  across all CLEF collections, faceted by Rocchio setting and relevance feedback size (i.e. value of $k$).}
	\label{fig:2}
\end{figure}

Finally, we study how relevance feedback effectiveness changes with the batch size (previously $k=25$). We use the BiolinkBERT retriever and vary $k$ from 5 to 50; results are shown in Figure~\ref{fig:2}. For AP, in general, the larger the feedback size the higher the effectiveness. Interestingly, we find that a small feedback size (e.g., $k=5$), which considers less feedback between query representation updates, is not more effective than larger ones.
This might be because relevant documents are rare in the assessment pool and thus smaller batches are more likely to contain only non-relevant documents. As in previous results, this reduces ranking effectiveness. Further, using larger feedback sizes is computationally beneficial as it requires fewer updates and re-rankings of the query representation.
We also observe that for AP, the best weight setting for the representations varies across feedback sizes, though most differences are not significant.
For Last Rel it appears that settings that put less importance on the feedback (and especially on the negative one) consistently yield higher effectiveness than other settings.

%% file: conclusions.tex
\section{Conclusion}

We considered the context of screening prioritisation for systematic review automation and adapted a generic relevance feedback mechanism that exploits dense retrieval. Unique to our settings is the fact that feedback is explicit and continuous, i.e. is provided iteratively as users screen documents. 
 Through extensive empirical experimentation, we reported that this method can achieve similar or better effectiveness in terms of AP and Last Rel compared to methods specifically designed for this task. In addition, it is computationally efficient as there is no need to re-train the ranker at each relevance feedback iteration, making it suitable for use in practice.